\documentclass[%
 reprint,
showpacs,
 amsmath,amssymb,
 aps,
]{revtex4-1}

\usepackage{graphicx}
\usepackage{dcolumn}
\usepackage{bm}

\begin{document}


\title{Collective neutrino oscillations and detectabilities in failed supernovae}

\author{Masamichi Zaizen$^{1}$}
\email{mzaizen@astron.s.u-tokyo.ac.jp}

\author{Takashi Yoshida$^{1}$}

\author{Kohsuke Sumiyoshi$^{2}$}

\author{Hideyuki Umeda$^{1}$}

\affiliation{%
$^{1}$Department of Astronomy, Graduate School of Science, University of Tokyo, Tokyo 113-0033, Japan\\
$^{2}$Numazu College of Technology, Ooka 3600, Numazu, Shizuoka 410-8501, Japan}%




\date{\today}

\begin{abstract}
	We investigate the collective neutrino oscillations under the three flavor multi-angle approximation in a spherically symmetric simulation of failed supernovae.
	A failed supernova emits high neutrino fluxes in a short time, while intense accretion proceeds with high electron density enough to experience re-collapse into a black hole.
	Our results show that matter-induced effects completely dominate over neutrino self-interaction effects and multi-angle matter suppression occurs at all time snapshots we studied.
	These facts suggest us that only MSW resonances affect the neutrino flavor conversions in failed supernovae and simple spectra will be observed at neutrino detectors.
	We also estimate the neutrino event rate in current and future neutrino detectors from a source at $10\mathrm{~kpc}$ as a Galactic event.
	The time evolution of neutrino detection could provide information about the dense and hot matter and constrain the neutrino mass ordering problem.
\end{abstract}

\maketitle


\section{Introduction}
Massive stars experience core collapse and end their lives.
When the central core reaches the nuclear matter density, the core bounces and produces a shock wave.
The shock wave gradually loses kinetic energy as it propagates through the envelope and then it stalls once.
In a delayed explosion scenario, hydrodynamical instabilities such as the standing accretion shock instability (SASI) and the neutrino-driven convection enhance the rate of the neutrino heating and then revive the stalled shock \cite{Janka16}.

Some fractions of the high mass progenitors fail to explode, called as failed supernovae (e.g., \cite{Sumiyoshi07, Sumiyoshi08}).
The numerical simulations of failed supernovae have been studied \cite{Sumiyoshi07, Sumiyoshi08, Fischer09, OConner11, Kuroda18}.
In these failed supernovae, the shock wave can not revive unlike successful supernovae due to the intense accretion of the outer layers.
As this shock wave can not blow the matter away, it falls into the proto-neutron star and the accretion contributes to continuously raise the temperature and the density in the center.
This is why more energetic neutrinos are emitted than in the successful supernovae.
This neutrino emission stops when the neutrino sphere is wrapped in the event horizon.
For only 1 second until re-collapse, $\sim 10^{58}$ neutrinos with each flavor are emitted from the proto-neutron star via the neutronization and the thermal pair production.
These neutrino spectra depend on an equation of state (EOS), so they enable us to obtain information about the EOS for the nuclear matter density \cite{Nakazato10}.

In general, high luminosity neutrinos lead to the neutrino self-interaction near the neutrino sphere \cite{Pantaleone95}.
This effect can cause the flavor transition as well as the matter-induced neutrino oscillations, known as the Mikheyev-Smirnov-Wolfenstein (MSW) effect \cite{MSW78, MSW85}.
This is called as collective neutrino oscillation and causes the spectral splits \cite{Fogli07} via synchronized oscillations \cite{Pastor02} and bipolar oscillations \cite{Hannestad06, Duan07}.
In consequence, the spectra of $\nu_e$ and $\bar{\nu}_e$ are swapped with the spectra of non-electron neutrinos $\nu_x$ and $\bar{\nu}_x$ at small radii in supernovae, where $\nu_x$ represents $\nu_{\mu}$ and $\nu_{\tau}$.

In order to study these phenomena, it is essential to use multi-angle approximation　(see Sec. II-A for detail), avoiding single-angle approximation which ignores dependence on the trajectories of neutrinos emitted from neutrino sphere \cite{Duan06a}.
Collective neutrino oscillations through high density matter have been explored in the previous papers \cite{Esteban08b, Chakraborty11a, Chakraborty11b}.
Multi-angle oscillations are completely suppressed when the matter density is dominant over the neutrino density ($n_{e^-}\gg n_{\nu}$).
In the comparable case ($n_{e^-}\sim n_{\nu}$), the {\it multi-angle decoherence} occurs due to the partial matter suppression.
These features can not be examined in the single-angle approximation.
As the neutrino self-interaction is a nonlinear effect, we have to numerically evaluate the multi-angle effect in realistic supernova simulations.
In addition, many studies based on a linear analysis have promoted a better understanding of the self-interaction effects (e.g. \cite{Sawyer09, Banerjee11}) and have revealed that the collective neutrino oscillations possess many instabilities besides the traditional oscillation phenomena which have ever been simulated.
Some of instabilities are expected to break the multi-angle matter suppression and also induce the flavor conversions near the core \cite{Raffelt13, Dasgupta15, Sawyer16}.

Recently, observations of failed supernovae have been making advance by the survey with the Large Binocular Telescope \cite{Kochanek08, Gerke15, Adams17a, Adams17b} and one candidate was identified.
Failed supernovae are thought to be unique sources which have information on a critical state before forming a black hole and provide keys to understand the formation scenario of stellar black holes.
As mentioned above, we will observe complicatedly swapped neutrino spectra of neutrino bursts.
Therefore it is important to investigate neutrino oscillation effects also in failed supernovae.

It is our purpose in this paper to investigate the three flavor collective neutrino oscillations in a 1D failed supernova model with the multi-angle approximation.
We evaluate the number of events of neutrinos from this model to be observed by current and future neutrino detectors.
In Sec. II, we introduce our method to calculate the neutrino oscillation and the  failed supernova model.
In Sec. III, we show the simulation results and detectability by the three detectors.
In Sec IV, conclusions are presented.
%
%
\section{Calculation method}
%
%
\subsection{Neutrino oscillation}
When we consider the neutrino oscillations including the collective effect, dense neutrino fluxes emitted from a neutrino sphere are described by density matrices $\rho$ and $\bar{\rho}$
\begin{eqnarray}
\rho_{\alpha\beta}(t,\mathbf{r},\mathbf{p}) = |\nu_{\alpha}\rangle \langle\nu_{\beta}|.
\end{eqnarray}
This diagonal component $\rho_{\alpha\alpha}$ is the number density of $\alpha$ flavor neutrinos.
The flavor evolution is given by solving the following von-Neumann equations \cite{Sigl93}
\begin{eqnarray}
i\left(\partial_t + \mathbf{v}\cdot\nabla_{\mathbf{r}}\right)\rho ~&&= \left[+H_{\mathrm{vac}}+H_{\mathrm{MSW}}+H_{\nu\nu},~\rho\right]   \label{eq.:EoM1} \\
i\left(\partial_t + \mathbf{v}\cdot\nabla_{\mathbf{r}}\right)\bar{\rho} ~&&= \left[-H_{\mathrm{vac}}+H_{\mathrm{MSW}}+H_{\nu\nu},~\bar{\rho}\right].   \label{eq.:EoM2}
\end{eqnarray}
This non-stationary term $\partial_t$ induces the temporal instability, which can enable the flavor conversion to occur at a small radius in a realistic supernova \cite{Abbar15, Dasgupta15}.
However, this requires us to treat both the flavor conversion and the neutrino transport and becomes more complicated.
In this paper, we assume the steady state, $\partial_t = 0$.
In Eqs.\eqref{eq.:EoM1}\eqref{eq.:EoM2}, $H_{\mathrm{vac}}$ is the Hamiltonian of vacuum oscillation and is expressed as
\begin{eqnarray}
H_{\mathrm{vac}} = \dfrac{1}{2E}U M^2 U^{\dagger},
\end{eqnarray}
where $E$ is the neutrino energy, $U$ is the Pontecorvo-Maki-Nakagawa-Sakata (PMNS) matrix \cite{PMNS1962} parametrized by the mixing angles $\theta_{ij}$, and $M^2$ is the squared mass matrix.
$H_\mathrm{MSW}$ is the potential induced by background charged leptons as
\begin{eqnarray}
H_{\mathrm{MSW}} = \sqrt{2}G_{\mathrm{F}}n_{e^{-}}\mathrm{diag}\left(1,0,0\right),
\end{eqnarray}
where $G_{\mathrm{F}}$ is the Fermi-coupling constant and $n_{e^{-}}$ is the electron number density as a function of the radius $r$.
In this potential, we ignore positrons and heavy charged leptons such as muon and tauon.
$H_{\nu\nu}$ is the Hamiltonian of the collective effect and is given by
\begin{eqnarray}
H_{\nu\nu} = \sqrt{2}G_{\mathrm{F}}\int\frac{\mathrm{d}^3q}{(2\pi)^3}\left(1-\cos\theta_{\mathbf{p}\mathbf{q}}\right)\left(\rho_{\mathbf{r},\mathbf{q}}-\bar{\rho}_{\mathbf{r},\mathbf{q}}\right),
\end{eqnarray}
where $\theta_{\mathbf{p}\mathbf{q}}$ denotes the interacting angle between two neutrinos with the momenta $\mathbf{p}$ and $\mathbf{q}$.
In order to calculate this quantity, we adopt the ``single-bulb model'' \cite{Duan06a} assuming that all neutrinos are uniformly and isotropically emitted from the neutrino sphere with radius $R_{\nu}$.
The radius of the neutrino sphere is assumed to be independent of neutrino energies and flavors, not the flavor dependent multi-bulb model which enhances the fast flavor conversions \cite{Chakraborty16b, Abbar18}.
We also neglect a neutrino halo effect \cite{Cherry12, Cherry13}.
This effect considers that a small fraction of neutrinos undergoes only a single direction-changing scattering outside the neutrino sphere and has a larger intersection angle.
Moreover, we impose the azimuthal symmetry on neutrino trajectories along the polar axis when calculating numerically.
This assumption has no influence on the inverted mass ordering case, but conceals unstable solutions in the normal mass ordering \cite{Raffelt13}.
More details will be explained in Sec. II-C.
Under these simplifications, we take account of the interactions of neutrinos from different angles.
This is multi-angle approximation.

In the three flavor case, the density matrix and the Hamiltonian can be expanded as a linear combination of the unit matrix and the Gell-Mann matrices. 
The coefficients of density matrices are called the polarization vector $\mathbf{P}$ and they are composed of 8 real components. 
Then we can transform the equations of motion into
\begin{eqnarray}
\frac{\mathrm{d}\mathbf{P}}{\mathrm{d}r}(E,u) &&= \left[\dfrac{+\omega(E)\mathbf{B}+\lambda\mathbf{L}}{v_{r,u}}+\mu\mathbf{D}(u)\right]\times\mathbf{P}(E,u) \\
\frac{\mathrm{d}\overline{\mathbf{P}}}{\mathrm{d}r}(E,u) &&= \left[\dfrac{-\omega(E)\mathbf{B}+\lambda\mathbf{L}}{v_{r,u}}+\mu\mathbf{D}(u)\right]\times\overline{\mathbf{P}}(E,u) \\
\mu\mathbf{D}(u) &&= \dfrac{\sqrt{2}G_{\mathrm{F}}}{2\pi R_{\nu}^2}\dfrac{R_{\nu}^2}{2r^2}\int\mathrm{d}E^{\prime}\mathrm{d}u^{\prime}\left(\dfrac{1}{v_{r,u}v_{r,u^{\prime}}}-1\right) \notag \\
&&~~\times\left(\mathbf{P}(E^{\prime},u^{\prime})-\overline{\mathbf{P}}(E^{\prime},u^{\prime})\right), \label{eq.:muD}
\end{eqnarray}
where $v_{r,u}$ and $u$ are defined as
\begin{eqnarray}
&&v_{r,u} = \sqrt{1-u\left(\frac{R_{\nu}}{r}\right)^2} \\
&&u = \sin^2\theta_R,
\end{eqnarray}
and the vectors $\mathbf{B}$ and $\mathbf{L}$ are the parameters corresponding to the mass term and the background electrons in the Hamiltonian, respectively, and are adopted from \cite{Dasgupta08a}.
Here $\theta_{R}$ is an emission angle relative to the radial direction at the neutrino sphere.
This transformation is in one-to-one correspondence to the intersection angle as $r\sin\theta = R_{\nu}\sin\theta_{R}$.

Here $\omega,\lambda,\mu$ are three kinds of oscillation frequencies and the cross product $\times$ is defined by the structure constants of the Gell-Mann matrices.
We define the strengths of vacuum oscillation, matter-neutrino interaction, and neutrino-neutrino interaction as
\begin{eqnarray}
\omega &&= \frac{|\Delta m^2_{31}|}{2E} \\
\lambda &&= \sqrt{2}G_{\mathrm{F}} n_{e^-} \\
\mu &&= \frac{\sqrt{2}G_{\mathrm{F}}}{4\pi r^2} \left(\frac{L_{{\nu}_e}}{\langle E_{{\nu}_e}\rangle} - \frac{L_{\bar{\nu}_e}}{\langle E_{\bar{\nu}_e}\rangle}\right),
\end{eqnarray}
where $\Delta m_{ij}^2 = m_i^2 -m_j^2$, $m_i$ is the eigenvalue of neutrino mass eigenstate $\nu_{i}$, $L_{\nu_{\alpha}}$ is the luminosity of $\nu_{\alpha}$, and $\langle E_{\nu_{\alpha}}\rangle$ is the average energy of $\nu_{\alpha}$.
Solar neutrino mass squared difference $\Delta m_{21}^2$ is included in the vector $\mathbf{B}$.
This $\mu$ is parametrized by ignoring the multi-angle term $(1-\cos\theta_{\mathbf{p}\mathbf{q}})$ in the neutrino self-interaction Hamiltonian and assuming that the $\nu_x$ flux is almost equal to the $\bar{\nu}_x$ one.

We also calculate the neutrino oscillations assuming the single-angle approximation to compare with the multi-angle ones.
Single-angle approximation ignores the angular dependence of the density matrices.
Then the angular integration of the multi-angle term in Eq.\eqref{eq.:muD} can be easily executed and be simplified into a geometric dilution factor $D(r)$.
We choose the representative angle as the radial direction $\theta_R = 0$ and then the self-interaction term is given by
\begin{eqnarray}
\mu\mathbf{D} &&= \frac{\sqrt{2}G_{\mathrm{F}}}{2\pi R_{\nu}^2}D(r)\int\mathrm{d}E^{\prime}\left(\mathbf{P}(E^{\prime})-\overline{\mathbf{P}}(E^{\prime})\right) \\
D(r) &&= \frac{1}{2}\left[1-\sqrt{1-\left(\frac{R_{\nu}}{r}\right)^2}~\right]^2.
\end{eqnarray}
This single-angle formalism is much easier to solve than the multi-angle calculations.

To obtain the correct multi-angle decoherence requires the large number of angle bins $N_u = \mathcal{O}(10^3)$.
The insufficient number of angular modes causes the decoherence of the collective neutrino oscillations at the small radius \cite{Duan06a, Esteban07}.
On the other hand, the energy resolution does not require a large number because the multi-angle decoherence is independent of the distribution of the neutrino energy.
However, we need to have the fine pattern of the spectral splits, so that we usually take the energy bins $N_E = \mathcal{O}(10^2)$.

In this work, we choose three flavor neutrino parameters as the following from \cite{PDG16}: 
$\Delta m_{21}^2 = 7.37\times 10^{-5}\mathrm{~eV^2}, ~\left|\Delta m^2_{31}\right|=2.46\times 10^{-3}\mathrm{~eV^2}, ~\sin^2 \theta_{12}=0.297, ~\sin^2 \theta_{13}=0.0218, \mathrm{~and~}\delta=0$.
We assume that CP violation effect $\delta$ can be neglected.
The mass ordering is normal if the sign of $\Delta m^2_{31}$ is positive and the inverted mass ordering is the negative case.

We treat the 3 flavor neutrino as the modified basis ($\nu_e$,$\nu_x$,$\nu_y$), not the ordinary flavor basis ($\nu_e$,$\nu_{\mu}$,$\nu_{\tau}$) \cite{Dighe2000, Dasgupta08a}.
This non-electron neutrino ($\nu_x$,$\nu_y$) is the rotated state by the neutrino mixing angle $\theta_{23}$;
\begin{eqnarray}
(\nu_e, \nu_x, \nu_y)^{\mathrm{T}} = R_{23}^{\dagger}(\theta_{23}) (\nu_e, \nu_{\mu}, \nu_{\tau})^{\mathrm{T}},
\end{eqnarray}
where $R_{23}$ is the rotation matrix.
The electron type neutrinos are not affected by this rotation.
%
%
\subsection{Failed supernova model}
We employ the spherically symmetric 1D failed supernova simulation \cite{Sumiyoshi07, Sumiyoshi08} with a progenitor having $40M_{\odot}$ \cite{Woosley95} and adopting the EOS by Lattimer \& Swesty with an incompressibility of $220\mathrm{MeV}$ (LS220-EOS) \cite{LS91}.

\begin{figure}
	\centering
	\includegraphics[width=0.8\linewidth]{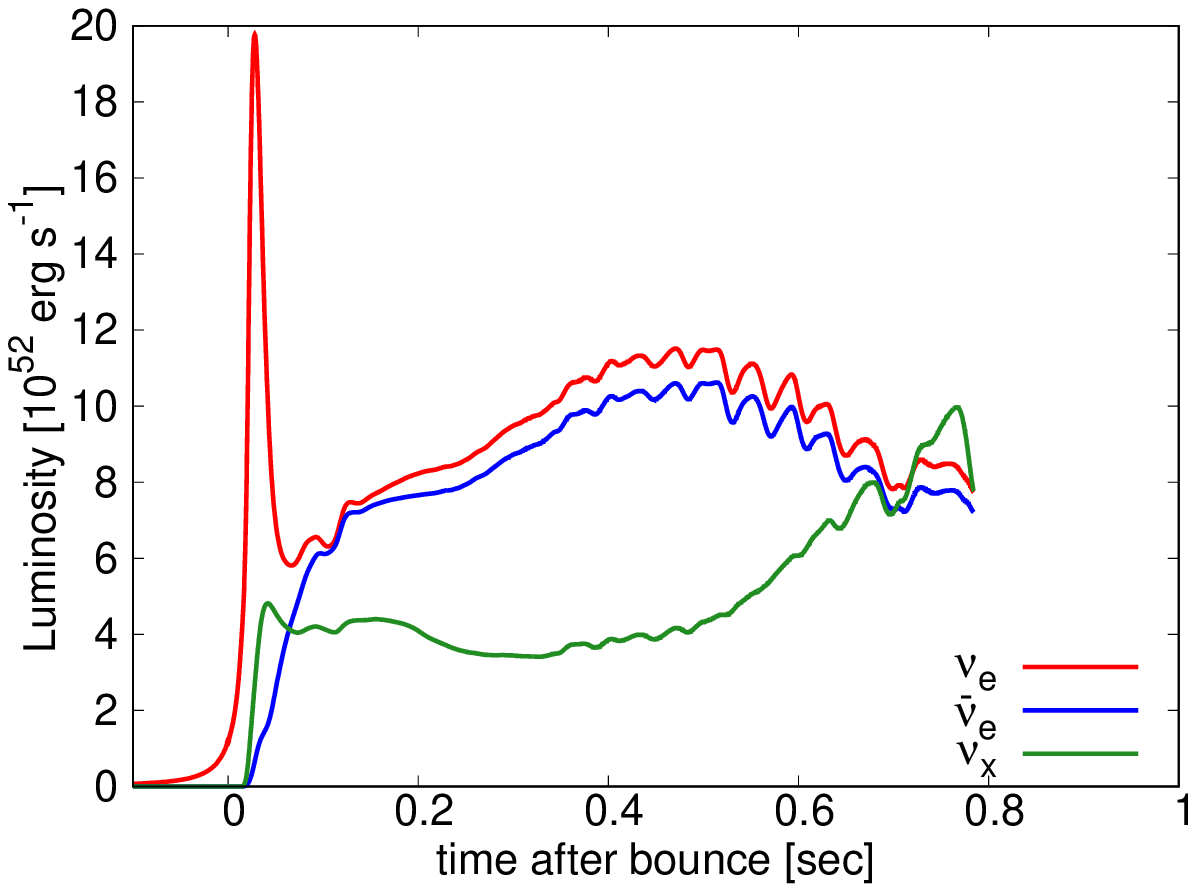}
	\includegraphics[width=0.8\linewidth]{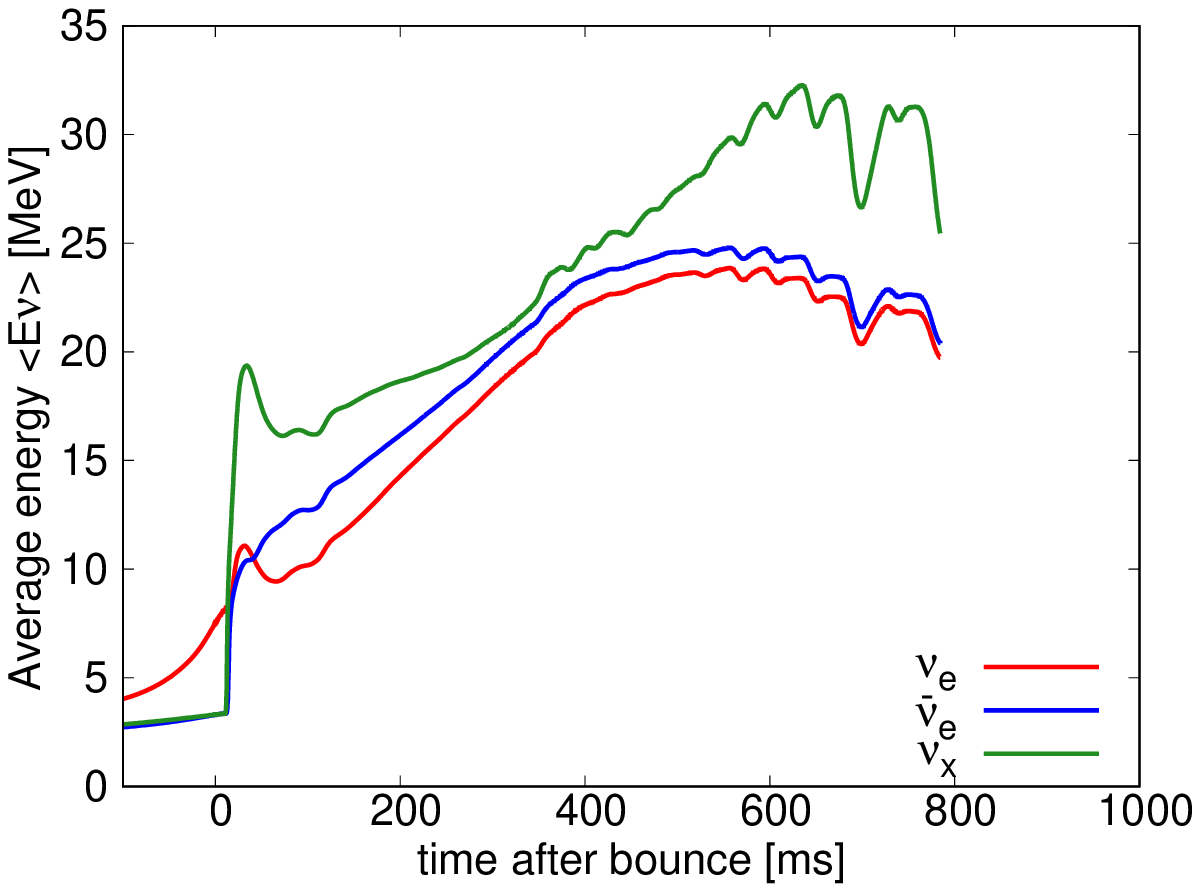}
	\caption{
		Luminosities (top) and averaged energies (bottom) of $\nu_{e}$(red line), $\overline{\nu}_{e}$(blue line), and $\nu_{x}$(green) as a function of $t_{\mathrm{pb}}$.
	}
	\label{fig:lumi_ave}
\end{figure}

Figure \ref{fig:lumi_ave} shows the luminosities $L_{\nu_{\alpha}}$ and the averaged energies $\langle E_{\nu_{\alpha}}\rangle$ of neutrinos as a function of time after bounce $t_{\mathrm{pb}}$ in this simulation.
As time passes, these two quantities increase because of the increasing temperature at the neutrino sphere.
A peak at $t_{\mathrm{pb}}=30\mathrm{~ms}$ in the luminosity of $\nu_e$ is due to the neutronization burst.
In this model, the proto-neutron star re-collapses into a black hole at the $t_{\mathrm{pb}}=783\mathrm{~ms}$ because the mass of proto-neutron star continues to increase.
Then the neutrino sphere is wrapped in the event horizon and the neutrino emission stops.

We use a Gamma distribution \cite{Tamborra12, Tamborra14} as an initial neutrino spectrum with $\alpha$ flavor at the surface of neutrino sphere:
\begin{eqnarray}
\varphi^{i}_{\nu_{\alpha}}(E_{\nu}) ~&&= \Phi(\nu_{\alpha})\dfrac{(E_{\nu})^{\xi_{\nu_{\alpha}}}}{\Gamma(\xi_{\nu_{\alpha}}+1)}\left(\dfrac{\xi_{\nu_{\alpha}}+1}{\langle E_{\nu_{\alpha}}\rangle}\right)^{\xi_{\nu_{\alpha}}+1} \notag \\
&&~~~~~~~~~~\times\exp\left[-(\xi_{\nu_{\alpha}}+1)\dfrac{E_{\nu}}{\langle E_{\nu_{\alpha}}\rangle}\right],
\end{eqnarray}
where $\Phi(\nu_{\alpha})=L_{\nu_{\alpha}}/\langle E_{\nu_{\alpha}}\rangle$ is a total number flux and $\xi_{\nu_{\alpha}}$ is a pinching parameter given by
\begin{eqnarray}
\xi_{\alpha} = \frac{\langle E_{\alpha}^2\rangle-2\langle E_{\alpha}\rangle^2}{\langle E_{\alpha}\rangle^2-\langle E_{\alpha}^2\rangle}.
\end{eqnarray}
These parameters are determined by the failed supernova simulation.
When we fit this pinching parameter, we adopt neutrino data below $100\mathrm{~MeV}$ .

We pick out $t_{\mathrm{pb}}=30, 100, 500, 600\mathrm{~ms}$ in time profile of this simulation and carry out post-process calculations in neutrino oscillations.
Table \ref{tab:SN_param} shows the parameters of neutrinos emitted from the neutrino sphere.
\begin{table*}
	\caption{Parameters of neutrinos on our failed supernova model at $t_{\mathrm{pb}}=30, 100, 500, 600\mathrm{~ms}$.
	}
	\centering
	\begin{tabular}{cccccccccccccc} 
		\tableline\tableline
		 & \multicolumn{4}{c}{Luminosity $(10^{52}~\mathrm{erg~s^{-1}})$} & \multicolumn{4}{c}{Averaged energy $\mathrm{(MeV)}$} & \multicolumn{4}{c}{Pinching parameter} & \multicolumn{1}{c}{neutrino sphere $\mathrm{(km)}$} \\
		 \cline{2-5} \cline{6-9} \cline{10-13} \cline{14-14}
		time & $L_{\nu_{e}}$ & $L_{\bar{\nu}_{e}}$ & $L_{\nu_{x}}$ & $L_{\bar{\nu}_{x}}$ & $\langle E_{\nu_{e}}\rangle$ & $\langle E_{\bar{\nu}_{e}}\rangle$ & $\langle E_{\nu_{x}}\rangle$ & $\langle E_{\bar{\nu}_{x}}\rangle$ & $\xi_{\nu_{e}}$ & $\xi_{\bar{\nu}_{e}}$ & $\xi_{\nu_{x}}$ & $\xi_{\bar{\nu}_{x}}$ & $R_{\nu}$ \\ 
		\tableline
		$30\mathrm{~ms}$  & $19.3$ & $0.861$ & $3.22$ & $3.25$ & $11.1$ & $10.2$ & $19.2$ & $19.3$ & $4.41$ & $3.75$ & $1.87$ & $1.89$ & $120$ \\ 
		$100\mathrm{~ms}$  & $6.37$ & $6.11$ & $4.14$ & $4.17$ & $10.2$ & $12.7$ & $16.2$ & $16.3$ & $3.27$ & $3.95$ & $1.85$ & $1.85$ & $110$ \\ 
		$500\mathrm{~ms}$  & $11.5$ & $10.6$ & $4.33$ & $4.34$ & $23.6$ & $24.6$ & $27.5$ & $27.5$ & $2.35$ & $2.61$ & $1.08$ & $1.08$ & $40$ \\ 
		$600\mathrm{~ms}$  & $10.3$ & $9.49$ & $6.07$ & $6.09$ & $23.6$ & $24.6$ & $31.1$ & $31.2$ & $2.27$ & $2.51$ & $1.17$ & $1.18$ & $40$  \\
		\tableline\tableline
	\end{tabular}
	\label{tab:SN_param}
\end{table*}

\begin{figure}
	\centering
	\includegraphics[width=0.8\linewidth]{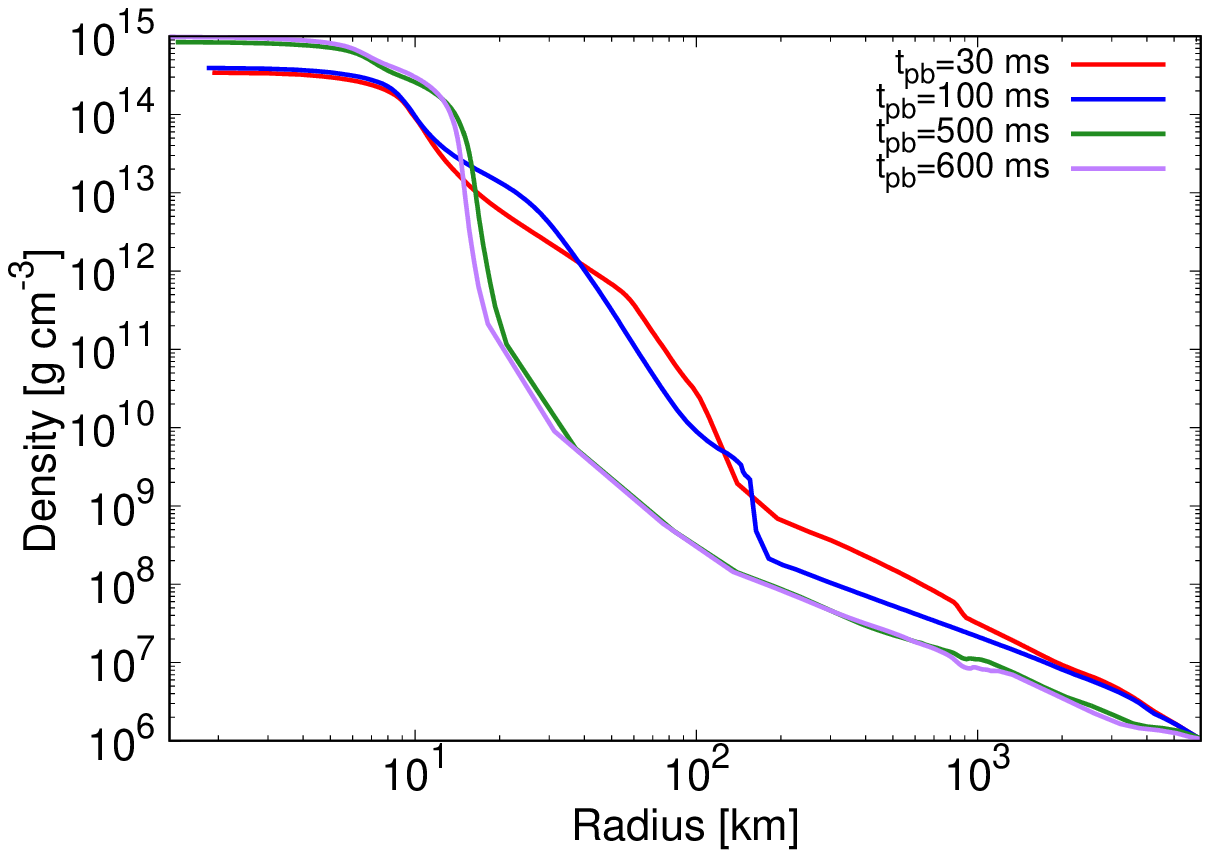}
	\caption{
		Density profiles at $t_{\mathrm{pb}}=30, 100, 500, 600\mathrm{~ms}$.
	}
	\label{fig:density}
\end{figure}

Figure \ref{fig:density} shows the density profiles at $t_{\mathrm{pb}}=30, 100, 500, 600\mathrm{~ms}$.
The shock wave is located at $r\sim 100\mathrm{~km}$ at earlier time $t_{\mathrm{pb}}=30, 100\mathrm{~ms}$, while it is near the proto-neutron star surface at the late time $t_{\mathrm{pb}}=500, 600\mathrm{~ms}$ and it forms the accretion shock.
The matter density in this failed supernova is higher than that in successful supernovae because of intense accretion and no shock revival.
Therefore we expect that the matter-induced effects have some impacts on the multi-angle collective effects.
In this study, we calculate the neutrino oscillations with the electron density profile of the failed supernova model in the multi-angle approximation as the standard case.
To confirm the matter-induced effects as well as the multi-angle collective effects, we also calculate additional two cases with different treatments of the electron density profile and angle approximation.
For the low density case, we reduce the electron density by a factor of $100$.
For the single-angle case, we change the treatment of the neutrino self-interaction term to the single-angle approximation.

\begin{figure}
\centering
\includegraphics[width=0.8\linewidth]{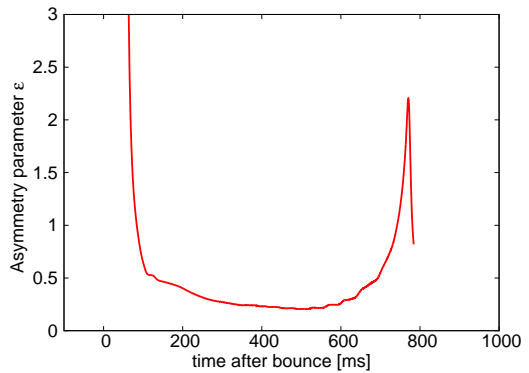}
\caption{
	The time evolution of the asymmetry parameter $\epsilon$.
}
\label{fig:asymmetry}
\end{figure}

We define a flavor asymmetry parameter between neutrinos and antineutrinos as
\begin{eqnarray}
\epsilon = \frac{\Phi(\nu_e) - \Phi(\nu_x)}{\Phi(\bar{\nu}_e) - \Phi(\bar{\nu}_x)} -1.
\end{eqnarray}
For a sufficiently small asymmetry parameter $\epsilon$, it is known that the multi-angle decoherence is triggered in both mass orderings \cite{Esteban07}.
Figure \ref{fig:asymmetry} shows the time evolution of the asymmetry parameter in the failed supernova model.
Asymmetry parameter always satisfies the condition $\epsilon \gtrsim 0.2$.
It is large enough to suppress the multi-angle decoherence and induce the quasi-single-angle behavior.
In this case, the multi-angle approximation should give similar feature to the single-angle one \cite{Esteban07}.
Consequently, the collective neutrino oscillation is not induced in the normal mass ordering.
%
%
\subsection{Linear analysis}
We mentioned the multi-angle collective neutrino oscillations do not occur in the normal mass ordering.
However, this is valid in the case that we assume the axial symmetry of neutrino propagation and ignore the the azimuthal dependence.
Recently the studies of linear analysis have reported that a different oscillation mechanism, the multi-azimuthal-angle (MAA) instability, would emerge without these assumptions \cite{Raffelt13, Mirizzi13, Chakraborty14, Chakraboty16a}.
This creates the unstable solutions which can break the axial symmetry and occurs only in the normal mass ordering.
The suppression of the MAA instability requires higher matter density than that of the traditional bimodal instability in the inverted mass ordering.
This has the possibility that the flavor conversions can occur even during the accretion phase of successful supernovae and requires detailed numerical simulations.
However, it is computationally hard to add azimuthal-angle bins $N_{\varphi}$ to energy and zenith-angle bins $N_E, N_{u}$, so we evaluate this instability by means of linearized analysis.
We　follow the linearized approach done in Ref. \cite{Raffelt13, Chakraboty16a}.

This failed supernova model does not have initial spectral pattern with the multiple spectral splits \cite{Dasgupta09} and the three-flavor effects would not emerge, so we simply use the two-flavor frameworks.
In order to see the evolution of an off-diagonal part of density matrix, it can be written by
\begin{eqnarray}
\rho = \frac{\mathrm{Tr}(\rho)}{2} + \frac{\Phi(\nu_e)-\Phi(\nu_x)}{2}g\begin{pmatrix}
1 & S \\ S^* & -1
\end{pmatrix},
\end{eqnarray}
where $g(E,u,\varphi)$ is the neutrino spectrum which does not depend on time and space.
We find the equation of motion in a stationary limit,
\begin{eqnarray}
i&&\left(v_r\partial_r + \mathbf{v}_T\cdot\mathbf{\nabla}_T\right)S  \notag \\
&&= \left[\omega +\lambda +\mu\int\frac{\mathrm{d}q^3}{(2\pi)^3}\kappa_{\mathbf{p}\mathbf{q}} g^{\prime}\right]S - \mu\int\frac{\mathrm{d}q^3}{(2\pi)^3}\kappa_{\mathbf{p}\mathbf{q}}g^{\prime}S^{\prime},  \notag \\
&&
\end{eqnarray}
where subscript $T$ denotes the spatial coordinates transverse to the radial coordinate $r$ and the factor $\kappa_{\mathbf{p}\mathbf{q}}$ is the multi-angle term $(1-\cos\theta_{\mathbf{p}\mathbf{q}})$.
Here, we can transform $S$ into the spatial Fourier mode,
\begin{eqnarray}
S(r,\mathbf{r}_T, E,u,\varphi) = \int \mathrm{d}k^2 \mathrm{e}^{-i\mathbf{k}\cdot \mathbf{r}_T}Q_{\mathbf{k}}\mathrm{e}^{-i\Omega_{\mathbf{k}}r}.
\end{eqnarray}
This inhomogeneous $\mathbf{k}$ mode brings up the spacial symmetry breaking for the assumed spherical symmetric model \cite{Mangano14, Duan15}.
However, we can sufficiently discuss whether instability region intersects with the supernova density profile, using only the largest-scale (homogeneous) mode $k=0$ \cite{Chakraboty16a}.
Therefore the linearized equation is given by
\begin{eqnarray}
\left[\frac{\omega +\bar{\lambda}}{v_r} - \Omega\right]Q = \frac{\mu}{v_r}\int\frac{\mathrm{d}q^3}{(2\pi)^3}\kappa_{\mathbf{p}\mathbf{q}}g^{\prime}Q^{\prime},
\end{eqnarray}
where $\bar{\lambda} = \lambda +\mu\int\frac{\mathrm{d}q^3}{(2\pi)^3}\kappa_{\mathbf{p}\mathbf{q}} g^{\prime}$ is an effective matter density.
This equation includes three instabilities: the bimodal instability in the inverted mass ordering, the multi-zenith-angle (MZA) instability and the MAA instability in the normal mass ordering.
The first two instabilities provide solutions with zenith-angle $u$ dependence in the inverted and normal mass ordering, respectively.
These solutions can be directly studied by numerical calculations.
Now we evaluate the MAA instability with nontrivial $\varphi$ dependence using this linearized equation.
%
%
\section{Results}
\subsection{Multi-angle matter suppression}
\begin{figure}
	\centering
	\includegraphics[width=0.8\linewidth]{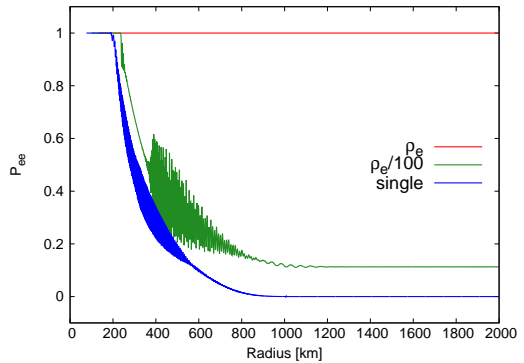}
	\caption{
		The radial evolution of $P_{ee}$ at $20\mathrm{~MeV}$ in the standard case (red solid line), the low density case (green solid line), and the single-angle case (blue solid line) at $t_{\mathrm{pb}}=600\mathrm{~ms}$.
	}
	\label{fig:t600_surv_IH}
\end{figure}

\begin{figure}
	\centering
	\includegraphics[width=0.8\linewidth]{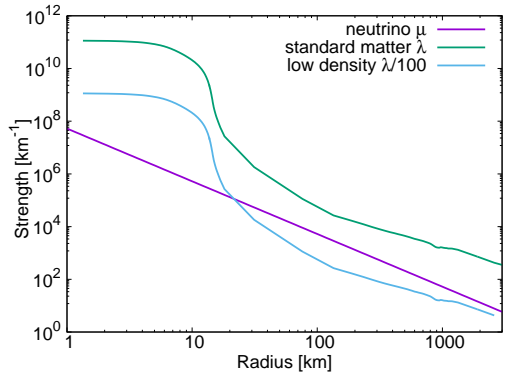}
	\caption{
		The strength of matter interaction $\lambda$ and neutrino self-interaction $\mu$ with radial coordinate at $t_{\mathrm{pb}} = 600\mathrm{~ms}$.
	}
	\label{fig:e_nu_t600}
\end{figure}

We first show the numerically calculated result in the inverted mass ordering at $t_{\mathrm{pb}}=600\mathrm{~ms}$ as an example.
Figure \ref{fig:t600_surv_IH} shows the radial evolution of the survival probability of electron neutrinos $P_{ee}$ at $20\mathrm{~MeV}$ in the standard case, the low density case, and the single-angle case.
For the standard case, we find that the collective neutrino oscillation is completely suppressed and the flavor conversion does not occur.
We do not see any flavor changes by the neutrino self-interactions in the all neutrino energy range.
Comparing the strength of the neutrino self-interaction with that of the matter effect, we find that the condition $\lambda(r) > \mu(r)$ is satisfied at all radii in Figure \ref{fig:e_nu_t600}.
This strong matter-induced effect prevents the instability of flavor conversions from growing.
By contrast, in the low density case, the collective neutrino oscillation is not suppressed and starts around $r\sim 200\mathrm{~km}$.
It is found that the neutrino self-interaction is comparable for the small matter effect at $r > 20\mathrm{~km}$ in Figure \ref{fig:e_nu_t600}.
The flavor instability is not affected by the matter suppression and brings about the quasi-single-angle behaviors.
Also, the result of the single-angle case is not affected by the matter suppression in Figure \ref{fig:t600_surv_IH}.
This means that the single-angle approximation enhances the flavor conversion in spite of high electron density.
In order to obtain physically reliable results of neutrino oscillation, we must adopt the multi-angle approximation including the matter suppression.

\begin{figure*}
	\centering
	\includegraphics[width=0.3\linewidth]{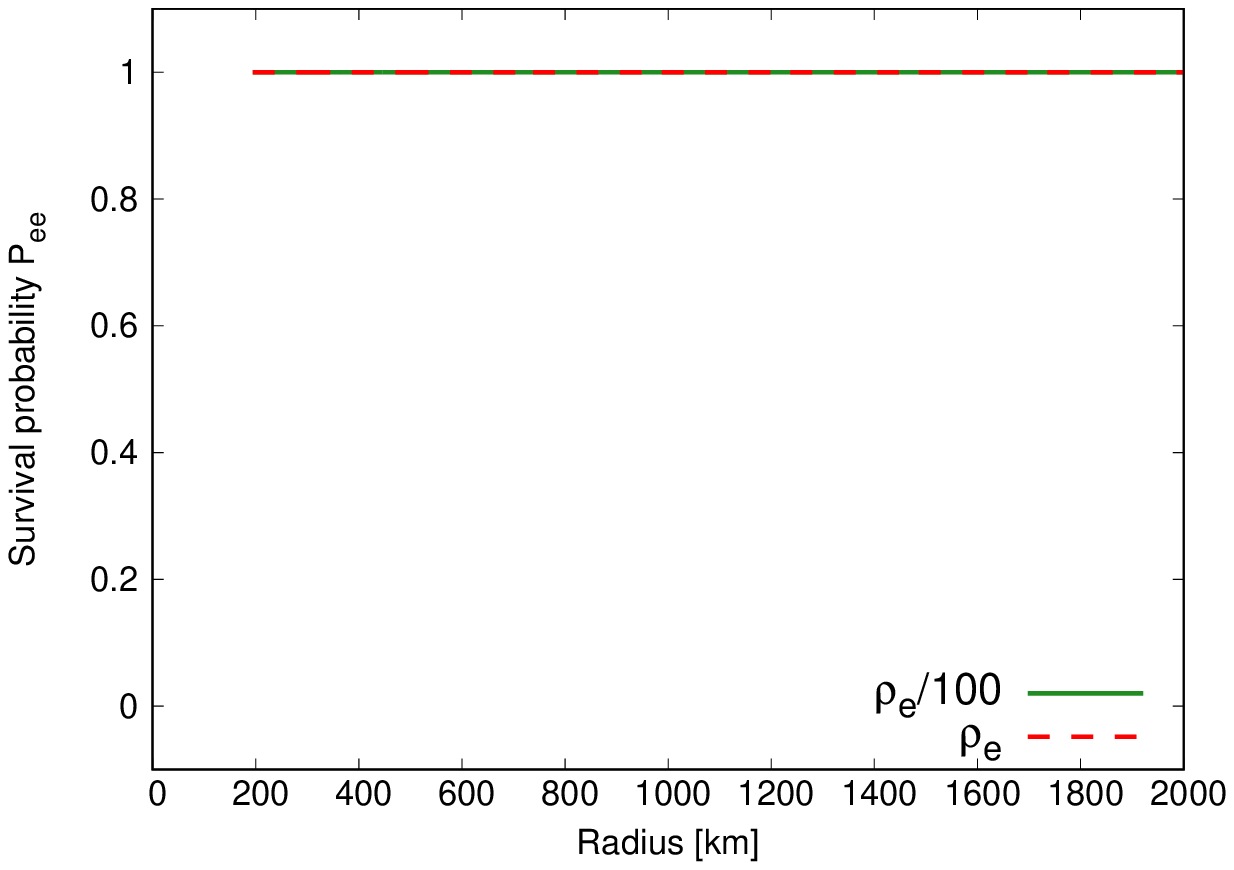}
	\includegraphics[width=0.3\linewidth]{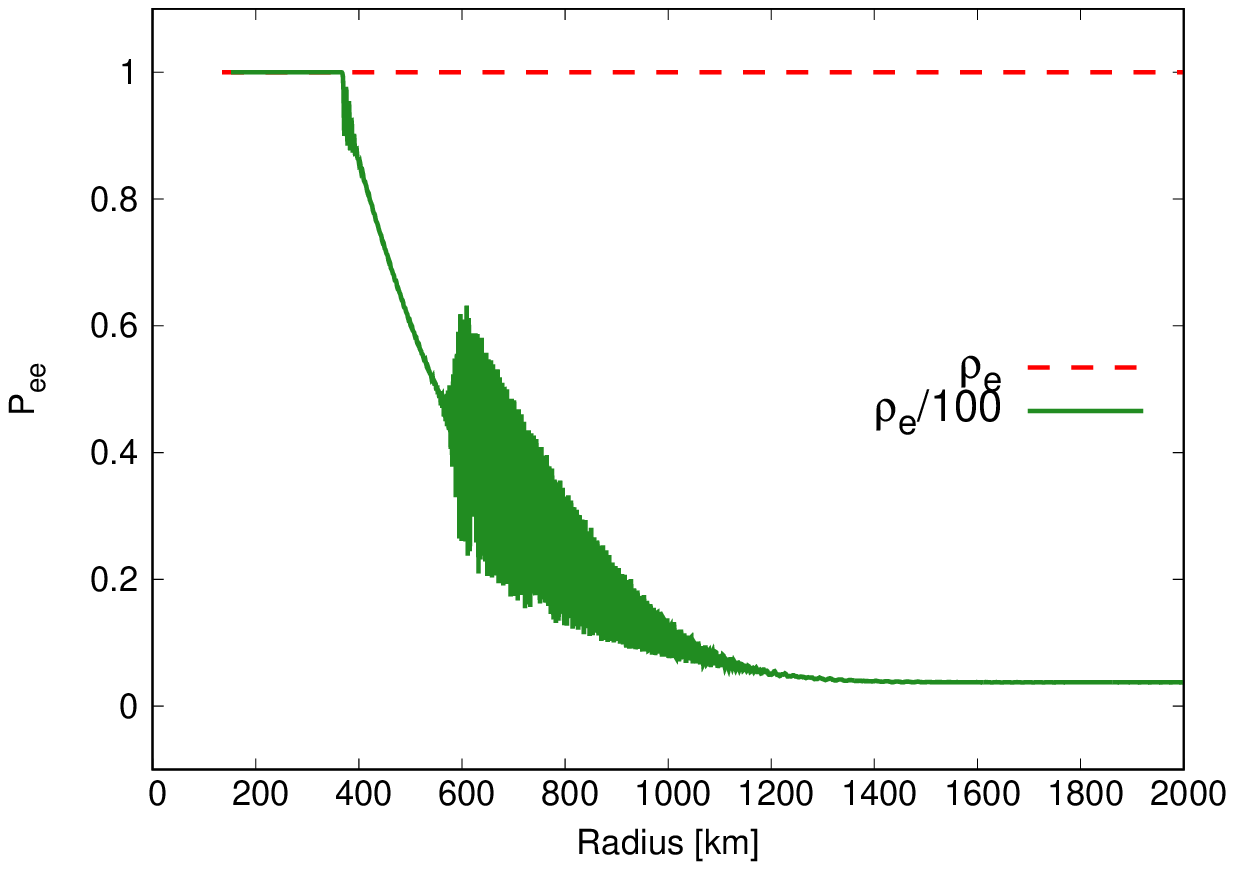}
	\includegraphics[width=0.3\linewidth]{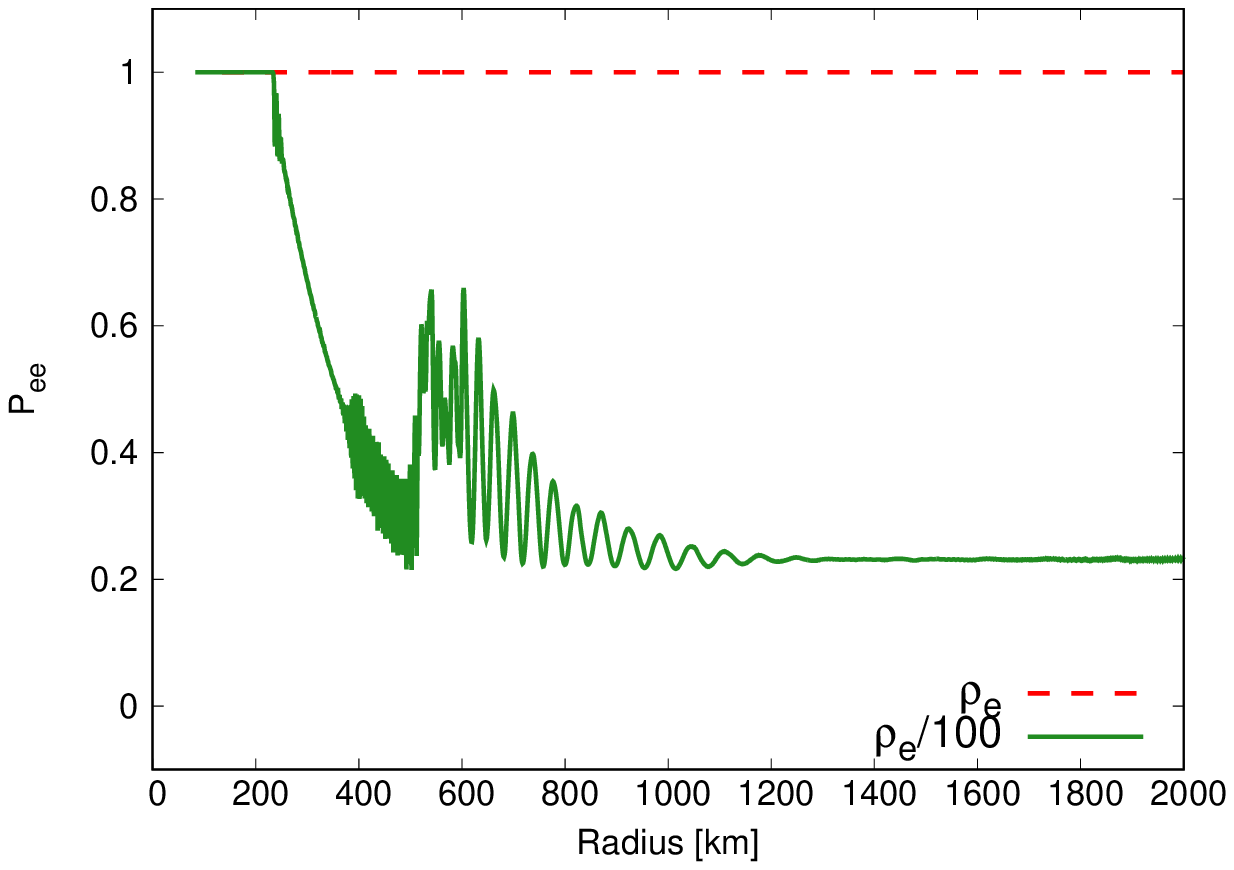}
	\caption{
		The radial evolution of $P_{ee}$ at $20\mathrm{~MeV}$ in the standard case (red solid line) and the low density case (green dotted line) at time $30\mathrm{~ms}$ (left), $100\mathrm{~ms}$ (middle), and $500\mathrm{~ms}$ (right).
	}
	\label{fig:surv_IH}
\end{figure*}

\begin{figure*}
	\centering
	\includegraphics[width=0.3\linewidth]{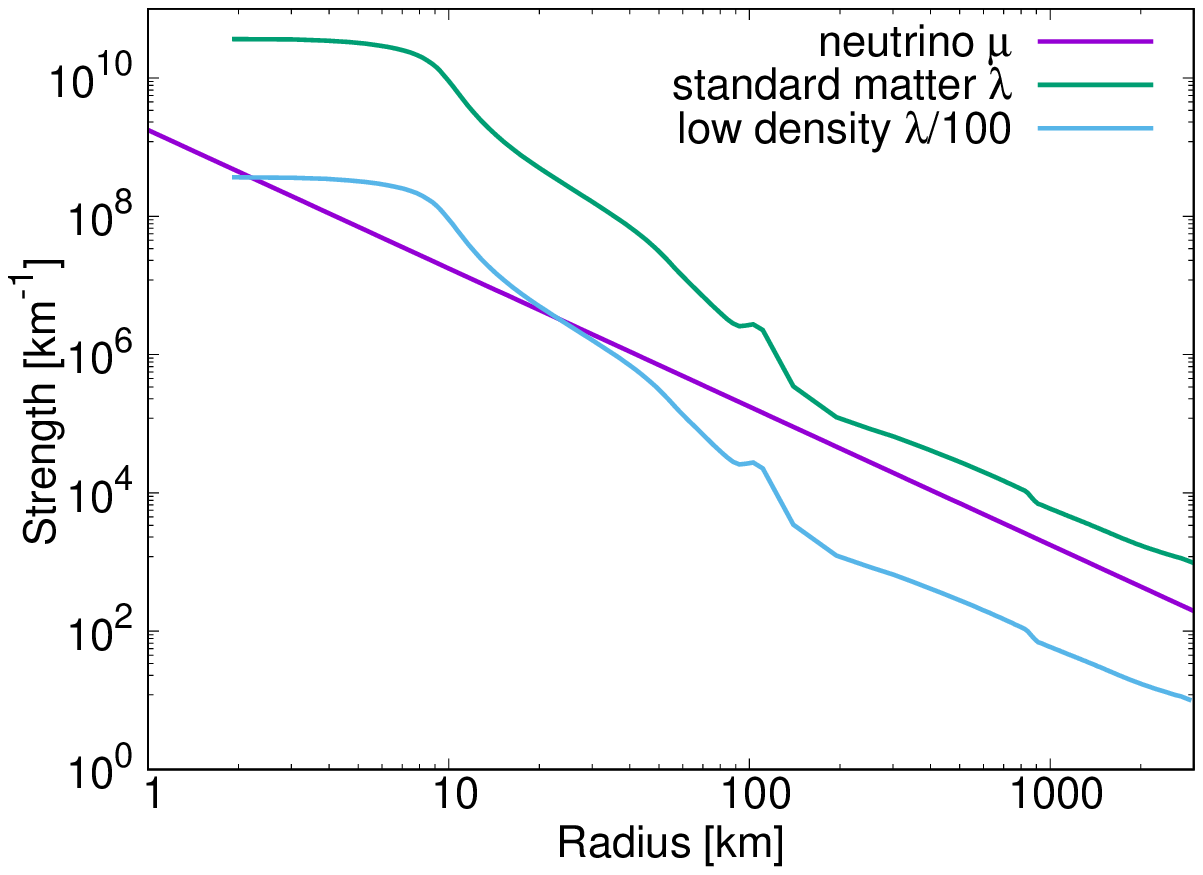}
	\includegraphics[width=0.3\linewidth]{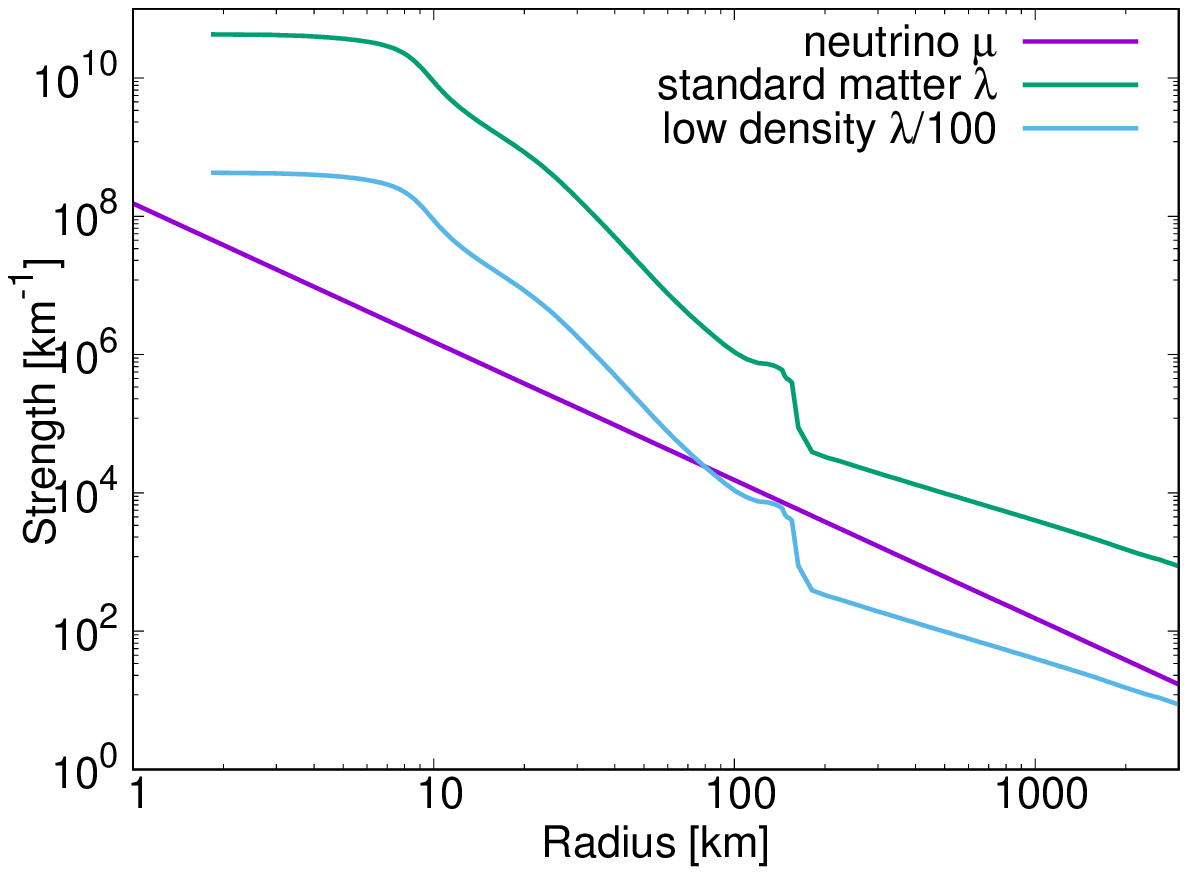}
	\includegraphics[width=0.3\linewidth]{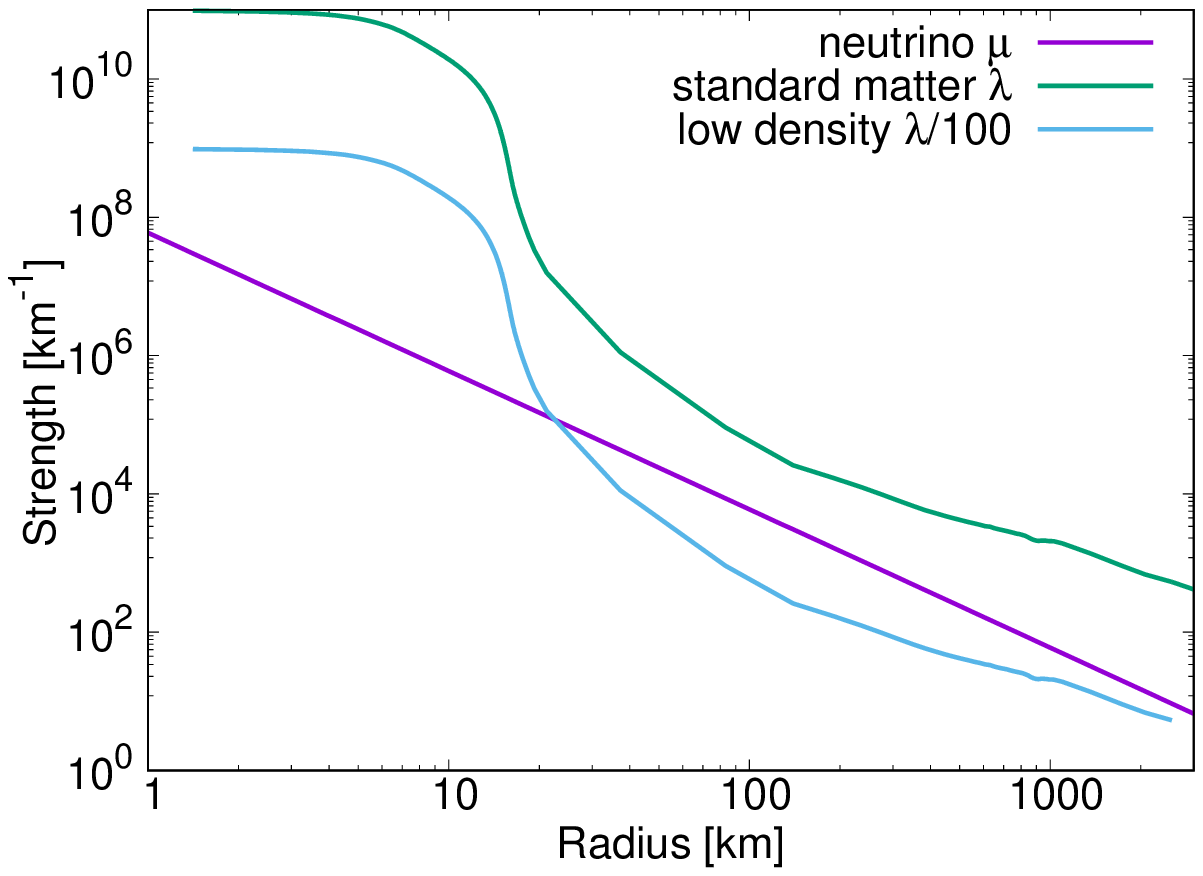}
	\caption{
		The strength of matter interaction $\lambda$ and neutrino self-interaction $\mu$ with radial coordinate at time $30\mathrm{~ms}$ (left), $100\mathrm{~ms}$ (middle), and $500\mathrm{~ms}$ (right).
	}
\label{fig:e_nu}
\end{figure*}

Figure \ref{fig:surv_IH} shows the radial evolution of the survival probability of electron neutrinos $P_{ee}$ at $20\mathrm{~MeV}$ at the other time steps, $t_{\mathrm{pb}}=30, 100, 500\mathrm{~ms}$.
We found that the multi-angle matter suppression dominates over the collective neutrino oscillation at all time steps.
In the low density case, the flavor conversion starts at $r\sim 400\mathrm{~km}$ for $t_{\mathrm{pb}}=100\mathrm{~ms}$.
The onset of the collective neutrino oscillation is delayed, compared with $t_{\mathrm{pb}}=500$ and $600 \mathrm{~ms}$.
Figure \ref{fig:e_nu} shows the strength of the neutrino self-interaction and the matter effect at the corresponding time steps.
As the shock wave is still located around $r\sim 100\mathrm{~km}$ at $t_{\mathrm{pb}}=100\mathrm{~ms}$, the radius satisfying $\mu\sim \lambda/100$ is large.
However it is small at $t_{\mathrm{pb}}=500$ and $600 \mathrm{~ms}$.
Therefore the development of the flavor instability is slow at $t_{\mathrm{pb}}=100\mathrm{~ms}$.

On the other hand, at $t_{\mathrm{pb}}=30\mathrm{~ms}$, when the neutronization burst occurs, the flavor transition can not be seen for both two density cases in Figure \ref{fig:surv_IH}.
In this phase, the excess of $\nu_e$ flux, $\Phi(\nu_e)\gg \Phi(\nu_x)\gg \Phi(\bar{\nu}_e)$, is achieved \cite{Kachelrie05}.
We can describe bipolar oscillations as the simultaneous pair conversion of $\nu_e \leftrightarrow \nu_x$ and $\bar{\nu}_e\leftrightarrow \bar{\nu}_x$.
This excess situation suppresses the bipolar conversions and causes only synchronized oscillations due to the large neutrino-antineutrino asymmetry \cite{Hannestad06}.
These synchronized oscillations are also suppressed by the high density.
Therefore, the collective neutrino oscillations do not occur at the neutronization burst.
%
%
\subsection{The MAA instability}
\begin{figure}
	\centering
	\includegraphics[width=0.8\linewidth]{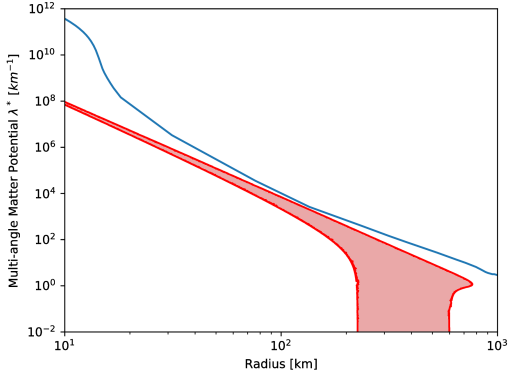}
	\caption{
		The unstable region where $\mathrm{Im}(\Omega)r > 1$ (red-shaded region) and the multi-angle matter potential $\lambda^*$ for the density profile (blue line) at $t_{\mathrm{pb}}=600\mathrm{~ms}$.
		The density profile does not intersect the MAA unstable regions at any radius.
	}
	\label{fig:t600_MAA}
\end{figure}

Next, we show the linearized analysis results in the normal mass ordering at $t_{\mathrm{pb}}=600\mathrm{~ms}$ as a representative.
Figure \ref{fig:t600_MAA} shows the density profile for our failed supernova model and the unstable regions with the growth rate, $\mathrm{Im}(\Omega)r > 1$, for the MAA instability.
For the linear analysis, we adopt the multi-angle matter potential as
\begin{eqnarray}
\lambda^* = \sqrt{2}G_{\mathrm{F}}n_{e^-}(r)\frac{R_{\nu}^2}{2 r^2} = \lambda\frac{R_{\nu}^2}{2 r^2},
\end{eqnarray}
where the factor $\frac{R_{\nu}^2}{2 r^2}$ arises from $v_r^{-1}$ in the linearized equation and this form is relevant for the stability check.

In this plot, the density profile does not intersect the MAA unstable regions at any radius.
The spatial small-scale flavor instabilities $k>0$ fill between the $k=0$ unstable region and the horizontal axis \cite{Chakraboty16a} in Figure \ref{fig:t600_MAA}.
So the large-$k$ instability modes are always stable and does not have any influence on the flavor conversions.
From this, we find that the MAA instability can not develop sufficiently and can be suppressed by the matter effect.

If we consider a broader neutrino halo flux, it gives additional unstable regions at larger radius for $\gtrsim 600\mathrm{~km}$ in Figure \ref{fig:t600_MAA} \cite{Sarikas12}.
However, the density profile is located above the original unstable regions and it does not also intersect the halo contribution.
Therefore, the suppression result allows us to ignore the neutrino halo effect.

We also confirm this suppression feature even at other time snapshots.
Therefore we conclude that the flavor conversions in the normal mass ordering are suppressed in our failed supernova model.
%
%
\subsection{Detection}
After the non-linear effect vanishes, the neutrino fluxes propagate through the MSW resonances.
The H resonance associated with $\Delta m_{31}^2$ and $\theta_{13}$ causes the complete conversion between $\bar{\nu}_e$ and $\bar{\nu}_y$ in the inverted mass ordering.
At this H resonance, the shock wave can influence the complete flavor conversion due to the sharp gradient of the density when it reaches the resonance region in the successful supernovae \cite{Kawagoe06}.
If the adiabatic condition \cite{Dighe2000}
\begin{eqnarray}
\gamma \equiv \frac{\Delta m^2}{2E}\frac{\sin^2 2\theta}{\cos 2\theta}\left[\frac{\mathrm{d}\ln n_e}{\mathrm{d}r}\right]^{-1} \gg 1
\end{eqnarray}
is violated in the H resonance region, the survival probability of anti-electron neutrinos becomes different from $\sin^2\theta_{13}\sim 0$ with each energy.

In general, we will observe the neutrino spectra which include the complicated combination of the flavor conversions such as the collective neutrino oscillation, the MSW resonance effect, the shock wave effect, and the Earth effect.
Therefore, it is difficult to extract information about the central region of supernovae from the observed spectra.

However as shown above, in failed supernovae the neutrino fluxes are not affected by the collective neutrino oscillation.
Moreover, strong shock wave can not propagate outward in the failed supernovae and the density of outer layer depends on the progenitor \cite{Woosley95}.
So the adiabaticity is always satisfied in the case of this model.
And the Earth effect was already discussed in \cite{Nakazato08} and it was mentioned that the accumulated event number has little dependence on the nadir angle.
Therefore if a failed supernova occurs at a near galaxy, we can obtain the neutrino spectra with only adiabatic MSW resonance effects through the envelope.
We evaluate this event rate, assuming that the distance to the target failed supernova is $d = 10 \mathrm{~kpc}$ as a typical distance in our Galaxy.
%
%
\subsubsection{Super-Kamiokande and Hyper-Kamiokande}
Super-Kamiokande is a water　Cherenkov detector in Japan.
In the fourth experiment (SK-IV), the fiducial volume is $22.5\mathrm{~kton}$ and the threshold energy is $3.5\mathrm{~MeV}$ corresponding to recoil electron kinetic energy \cite{Sekiya13, Sekiya16}.
Super-Kamiokande has a high sensitivity to inverse beta decay (IBD; $p(\bar{\nu}_e,e^+)n$) reaction and electron anti-neutrinos can be detected via this reaction.
The threshold energy for $\bar{\nu}_e$ is $E_\mathrm{th} = 4.79\mathrm{~MeV}$ because of the mass difference between proton and neutron.
We take the cross section $\sigma$ of IBD reaction from \cite{Strumia03}.

Hyper-Kamiokande is designed as the next generation experiment with the water Cherenkov detector \cite{HK_design_2018}.
The fiducial volume is $380\mathrm{~kton}$ and $16.9$ times as large as that in Super-Kamiokande.
Assuming that Hyper-Kamiokande has the same threshold energy $4.79\mathrm{~MeV}$ for $\bar{\nu}_e$, the detected event rate should be also $16.9$ times of SK-IV.

The event rate detected by Super-Kamiokande and Hyper-Kamiokande is estimated by
\begin{eqnarray}
\dfrac{\mathrm{d}N_{\bar{\nu}_e}}{\mathrm{d} t} = \dfrac{N_{p}}{4\pi d^2}\int^{\infty}_{E_{\mathrm{th}}}\mathrm{d}E~ \varphi(E)\sigma(E),
\end{eqnarray}
where $N_p$ is the number of target protons in the water tank, $d$ is the distance to a failed supernova and $\varphi$ is the spectrum of electron antineutrino at the detector.

\begin{figure}
	\centering
	\includegraphics[width=0.8\linewidth]{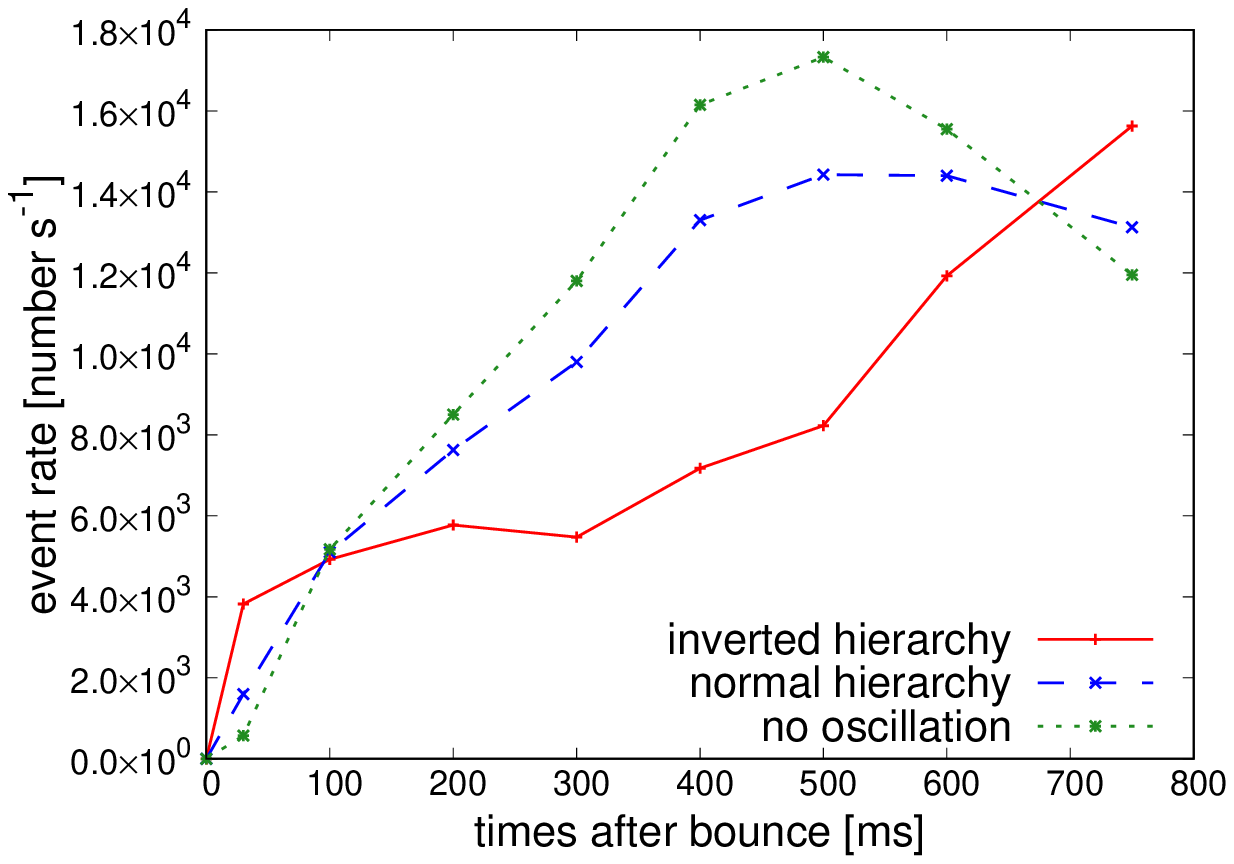}
	\includegraphics[width=0.8\linewidth]{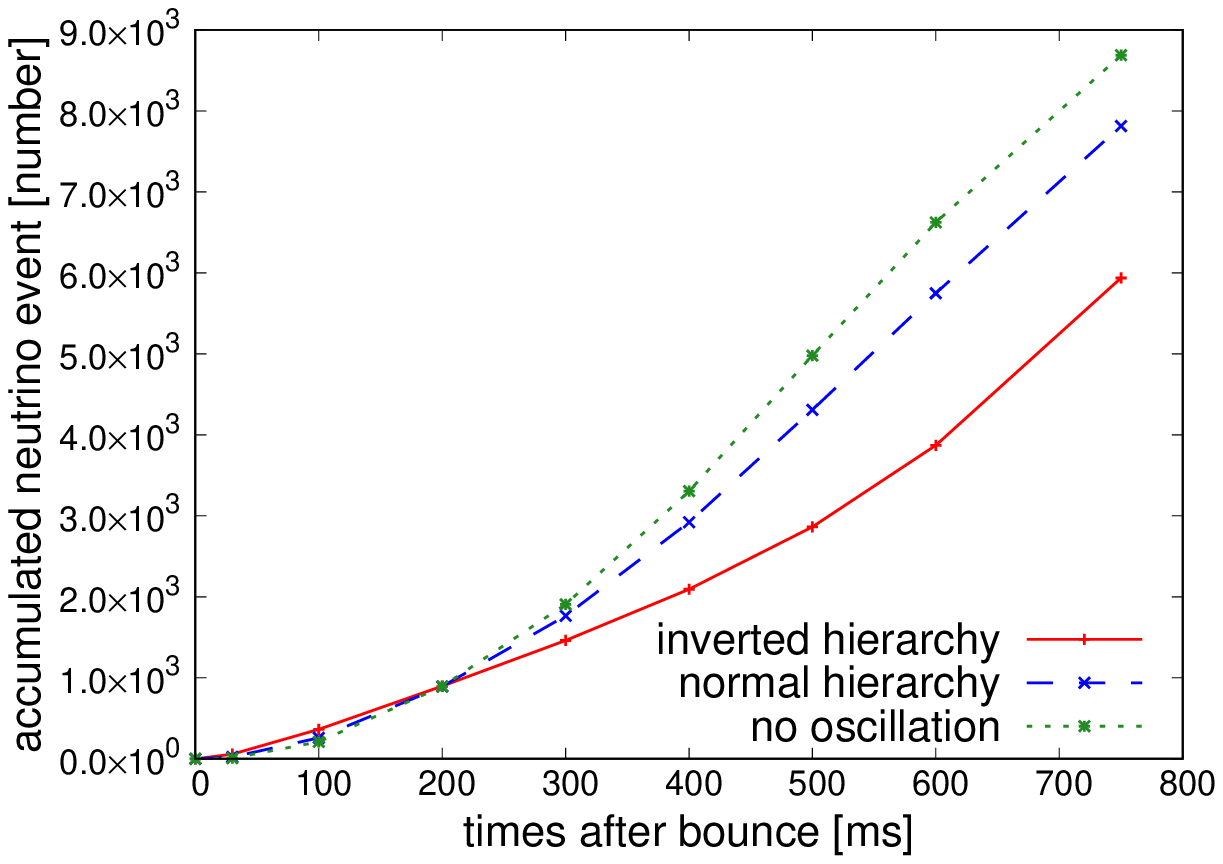}
	\caption{
		The event rate (top) and accumulated event number (bottom) at Super-Kamiokande from a failed supernova at $d=10\mathrm{~kpc}$. The inverted mass ordering, the normal mass ordering, and no oscillation case are shown by red solid, blue dashed,  and green dotted lines.
	}
	\label{fig:SK_det}
\end{figure}

\begin{figure}
\centering
\includegraphics[width=0.8\linewidth]{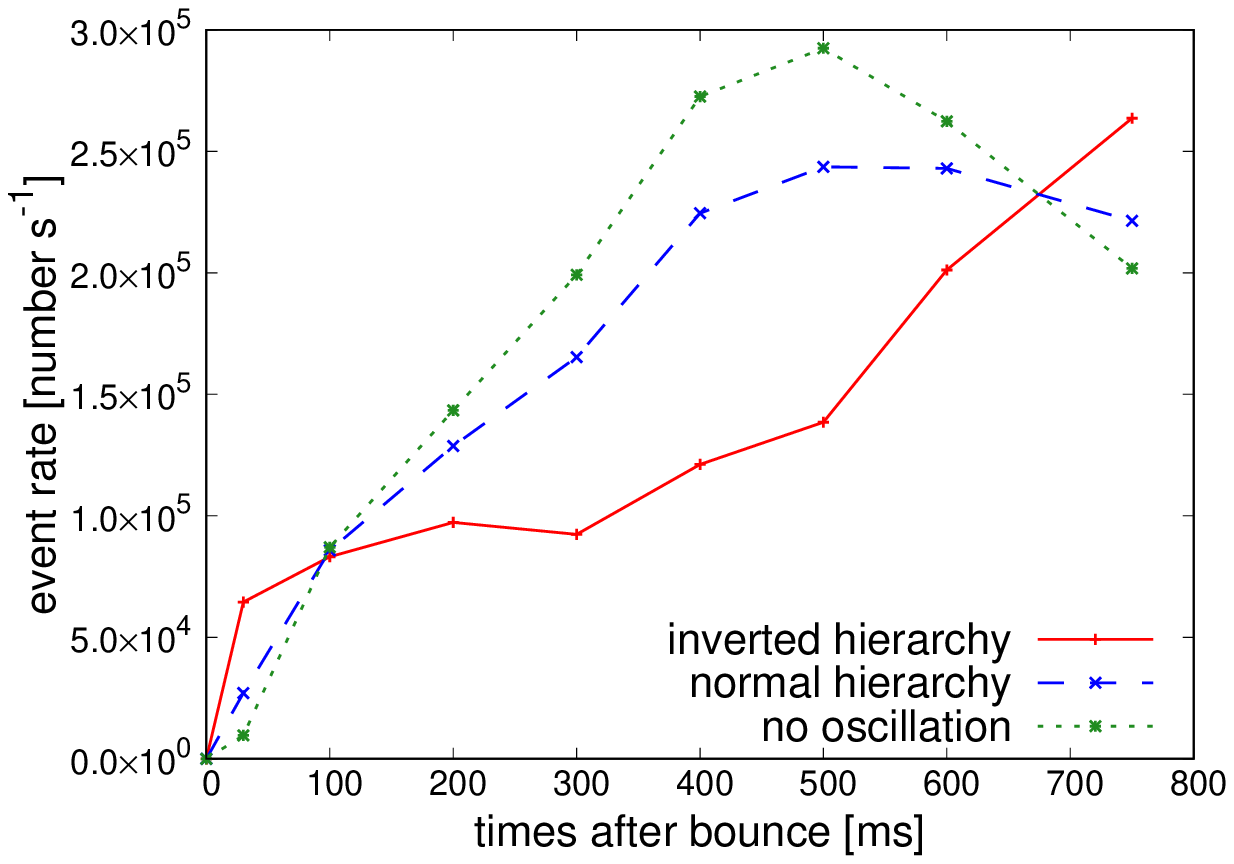}
\includegraphics[width=0.8\linewidth]{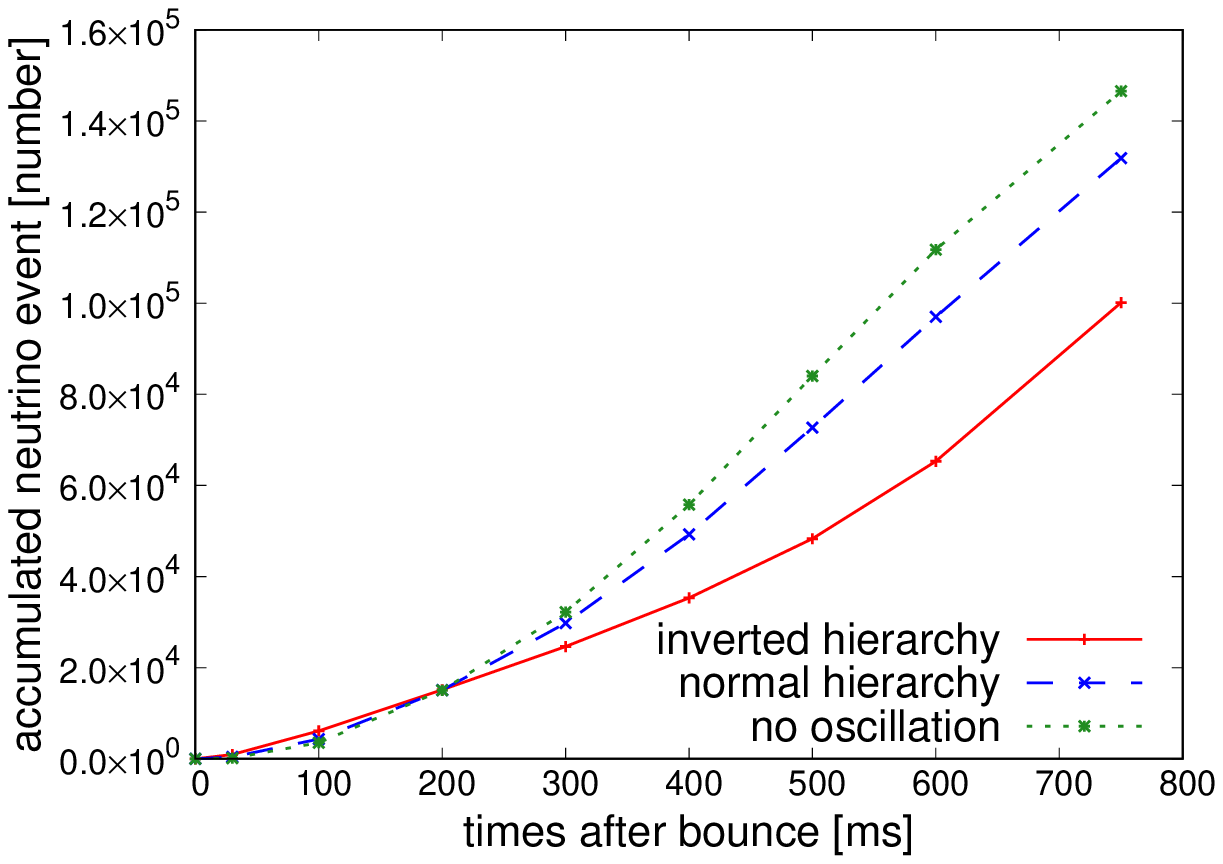}
\caption{
	The event rate (top) and accumulated event number (bottom) at Hyper-Kamiokande from a failed supernova at $d=10\mathrm{~kpc}$. The inverted mass ordering, the normal mass ordering, and no oscillation case are shown by red solid, blue dashed,  and green dotted lines.
}
\label{fig:HK_det}
\end{figure}

Figures \ref{fig:SK_det} and \ref{fig:HK_det} show the event number rate and the accumulated detection number at Super-Kamiokande and Hyper-Kamiokande, respectively.
There are big differences between the inverted and normal mass orderings in the neutrino detection rate.
Especially, the accumulated neutrino event number in the normal mass ordering is $1.5$ times (about $2000$, $30000$ events) larger than in the inverted mass ordering at stopped time $t\sim 750\mathrm{~ms}$ at Super-Kamiokande and Hyper-Kamiokande, respectively.
Since the adiabatic H resonance affects only antineutrinos as complete conversion $\bar{\nu}_e\leftrightarrow \bar{\nu}_y$ in the inverted mass ordering, the spectrum of $\bar{\nu}_e$ observed by Super-Kamiokande is transformed into the initial spectrum of heavy neutrinos $\bar{\nu}_{\mu}$ and $\bar{\nu}_{\tau}$.
This means that the event rate in the inverted mass ordering can be regarded as the evolution of the number flux of non-electron neutrinos.
On the other hand, the H resonance unchanges antineutrino flavors in the normal mass ordering according to the level crossing scheme.
Therefore the mixed spectra with $\bar{\nu}_e$ and $\bar{\nu}_x$ by the L resonance will be detected.
These features produce the result that the accumulated number event in the inverted mass ordering is smaller than that in the normal mass ordering.

The event rate in the normal mass ordering has a peak at $t_{\mathrm{pb}}=500\mathrm{~ms}$, while that in the inverted mass ordering continuously increases until the black hole formation.
Since the observed spectrum in the normal mass ordering includes $\bar{\nu}_e$, the event rate decreases at the late time.
In our model, Figure \ref{fig:lumi_ave}, the increase of luminosity and averaged energy of $\bar{\nu}_e$ ceases after $500\mathrm{~ms}$, whereas those of $\bar{\nu}_x$ continue to increase.
The neutrino spheres of $\nu_e$ and $\bar{\nu}_e$ are determined by both the charged-current and neutral-current reactions on nucleons.
On the other hand, $\nu_x$ interacts only via the neutral-current reaction.
Therefore the neutrino spheres of $\nu_x$ is located in the innermost radius.
An increase of temperature at this non-electron neutrino sphere due to the contraction of proto-neutron star causes the increase of non-electron neutrino pair production rate via thermalized electron-positron pairs, $e^- +e^+ \to \nu_{x}+\bar{\nu}_x$ \cite{Fischer09}.
This leads to the drastic increase of the luminosity of $\nu_x$.
%
%
\subsubsection{DUNE}
The Deep Underground Neutrino Experiment (DUNE) is a proposed neutrino experiment in the United States \cite{DUNE_design}.
A Liquid Argon Time-Projection Chamber can provide a detection of electron neutrino via a charged-current reaction $^{40}\mathrm{Ar}(\nu_e,e^-) ^{40}\mathrm{K}^*$.
DUNE will be composed of four detectors with the fiducial volume $10\mathrm{~kton}$.
Therefore the total fiducial volume of liquid argon is designed to be $40\mathrm{~kton}$.
The threshold energy of this charged-current reaction has not been determined and, here, we assume that the electron energy cut-off is $5\mathrm{~MeV}$.
Due to the energy difference between $^{40}\mathrm{Ar}$ and $^{40}\mathrm{K}^{*}$, the threshold energy for electron neutrinos is $8.28\mathrm{~MeV}$.
We take the cross section of $^{40}\mathrm{Ar}(\nu_e,e^-) ^{40}\mathrm{K}^*$ from \cite{Suzuki13}.

\begin{figure}[b]
	\centering
	\includegraphics[width=0.8\linewidth]{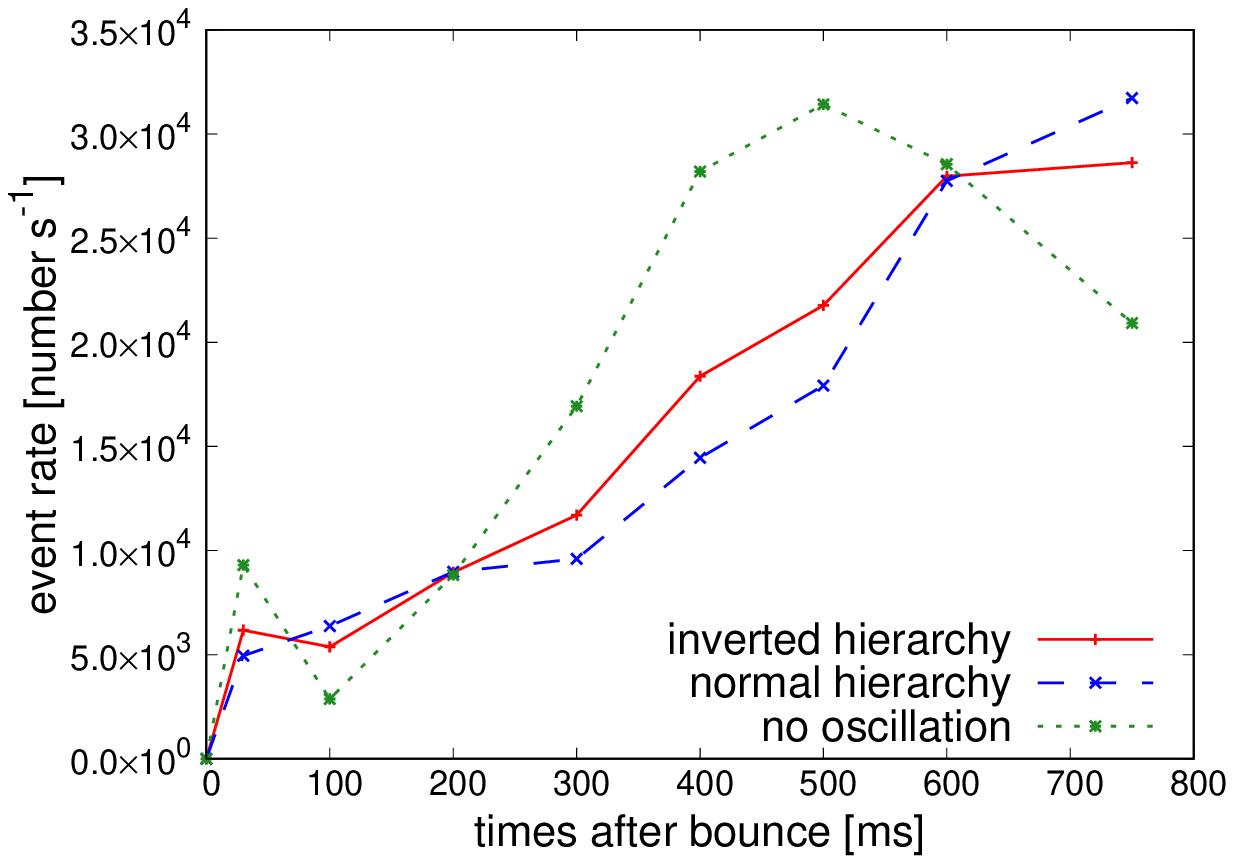}
	\includegraphics[width=0.8\linewidth]{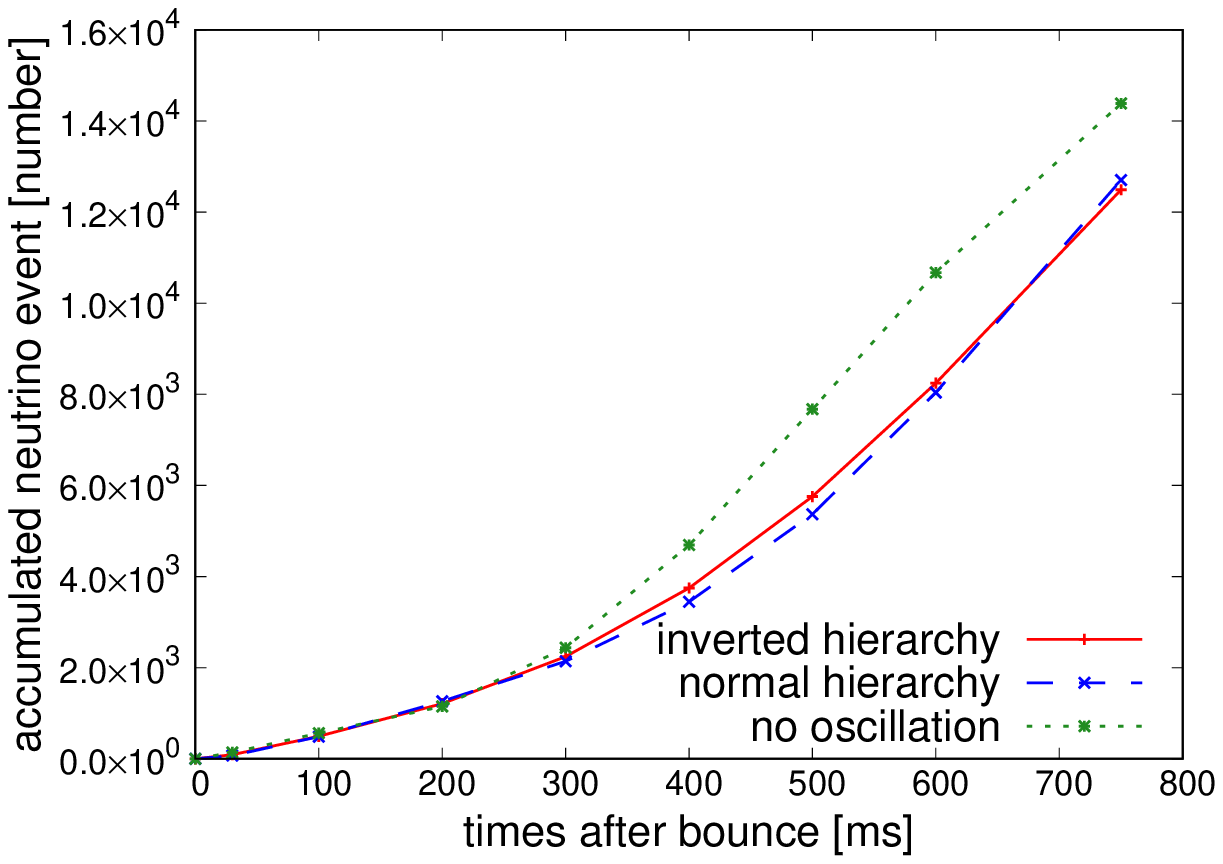}
	\caption{
		The event rate (top) and accumulated event number (bottom) at DUNE from a failed supernova at $d=10\mathrm{~kpc}$. The inverted mass ordering, the normal mass ordering, and no oscillation case are shown by red solid, blue dashed,  and green dotted lines.
	}
	\label{fig:DUNE_det}
\end{figure}
Figure \ref{fig:DUNE_det} shows the expected electron neutrino event number by DUNE.
It is found that the small peak at $t_{\mathrm{pb}}=30\mathrm{~ms}$ does not exist in the event rate in the normal mass ordering.
In the inverted mass ordering, the H resonance has little impact on spectrum of $\nu_e$ and the L resonance blends $\nu_e$ and $\nu_y$, while the complete conversion $\nu_e\leftrightarrow \nu_y$ occurs at the H resonance in the normal mass ordering and the features of $\nu_e$ spectrum are completely lost.
Therefore if the peak of the neutronization burst is observed in the event rate, it means that the mass ordering is inverted.

No significant difference can be seen for the accumulated neutrino event numbers in both mass orderings.
This is because the cross sections via charged-current reaction are more sensitive to high energy neutrinos than those via IBD reaction in water Cherenkov detectors.
Higher energy components than the peak energy of neutrino spectrum become dominant in the event rate, but the tail of neutrino spectra has smaller difference in flavors than around the peak.
So the mixed spectra by the MSW resonances are almost identical between both mass orderings.

We see different features in the time variation of the neutrino event rate between the normal and inverted mass orderings.
In the normal mass ordering, the event rate continuously increases until the final stage, whereas the rate of increase becomes slow in the inverted mass ordering.
This is the opposite feature to the event rate in Hyper-Kamiokande.
This is because the H resonance  completely converts the spectrum of $\nu_e$ into that of $\nu_y$ in the normal mass ordering.
The event rate does not include $\nu_e$ fluxes, and therefore it continue to increase until the black hole formation in the normal mass ordering.

Also, the determination of the mass ordering using the IceCUBE neutrino observatory has been proposed by \cite{Abbasi11, Serpico12}.
IceCUBE can detect the Cherenkov light via neutrino reactions such as electron scattering and IBD reaction with the $2\mathrm{~ms}$ time binning and this is short enough to resolve the rise time at the neutronization burst $\mathcal{O} = (10)\mathrm{~ms}$ for a galactic supernova.
By comparing IceCUBE and other neutrino detectors with Super-Kamiokande and DUNE, the stronger restrictions on this problem will be obtained.

We note that this tendency of neutrino emission depends on the progenitor structure and the mass accretion rate at the neutrino sphere of the proto-neutron stars.
In the case of failed supernovae with large mass accretion rates, corresponding to our model, the time evolution of the neutrino spectra will be similar to the one obtained in this study \cite{Sumiyoshi08, Fischer09, Kuroda18}.
We would observe the time evolution of the event rate similar to the one expected above.
On the other hand, failed supernovae with small mass accretion rate show the decrease of luminosities for all flavor neutrinos at late time \cite{Fischer09}.
Since the luminosity of $\nu_x$ expresses the similar behavior to that of $\nu_e$, the differences between the neutrino mass orderings would not be seen in the event rate.
%
%
\section{Conclusions}
We studied the three flavor multi-angle collective neutrino oscillation using a 1D failed supernova model of a $40 M_{\odot}$ progenitor with LS220-EOS.
We showed the time evolution of the number count for a current and future detectors, assuming that the failed supernova is located at $10\mathrm{~kpc}$ as a Galactic event.

We obtained the results that the collective neutrino oscillations in the failed supernova model were completely suppressed by the matter-induced effect at all time snapshots from the neutronization burst to the formation of a black hole.
More neutrinos with higher energy are emitted from failed supernovae than those from successful supernovae.
This makes the neutrino-neutrino interaction stronger.
However, the accretion materials which can not be blown away by the shock wave also make the electron density higher.
This massive matter-induced effect prevents the collective flavor instability from growing.
We will observe the neutrino spectra simply affected by only adiabatic MSW resonances, unlike successful supernovae.
Complete matter suppression allows us easily understand the detected spectra in the neutrino detector.

On the other hand, in the single-angle case, this matter suppression was not be seen at all.
Neutrinos propagating through different trajectories to the same location experience larger matter effects than that through radial direction.
This angular propagation leads to the multi-angle matter suppression.
In the single-angle approximation, the matter suppression does not occur and the single-angle approximation produces physically unreliable neutrino oscillations.
When lower electron density than the original one is adopted, the quasi-single-angle oscillations with partial matter suppressions were seen.
This means that we can not consider the neutrino-neutrino interaction separately from hydrodynamical model.
Thus we must carry out the detailed multi-angle calculations and should not neglect the matter-induced effects in a realistic failed supernova model.

Our evaluations in the neutrino event rate also indicates the possibility to determine the neutrino mass ordering problem.
The neutronization burst would be a important tool to resolve the mass ordering problem if the neutrino fluxes at the Earth is large enough for DUNE to resolve the neutrino incident counts in time bins.
The comparison of the time evolution of the event rate at Hyper-Kamiokande with that at DUNE could also present the neutrino mass ordering to us, when the mass accretion to the proto-neutron star in a failed supernova is large.
If the time evolution of the event rate at Hyper-Kamiokande increases continuously and that at DUNE weakens at the late time, they would mean that the mass ordering is inverted.
In the opposite case, they would teach us the normal mass ordering.
Therefore the simultaneous observation in $\nu_e$ and $\bar{\nu}_e$ is important.

In this paper, we discussed the neutrino detection from a failed supernova using the traditional multi-angle simulation and the MAA instability.
On the other hand, many studies on a linear analysis have also revealed the existence of other instabilities which the the self-interaction effects produce.
These instabilities break the assumed symmetries spontaneously and have possibility that provides us the complicated flavor conversions.
Estimating the detected event number and the observed spectra from a realistic supernova requires us to tackle with these obstacles numerically.
However, many of these have not been simulated due to computational difficulties yet and this is further study.

\begin{acknowledgments}
This work has been partly supported by Grant-in-Aid for Scientific Research on Innovative Areas (26104006, 26104007, 17H06357, 17H06365) from the Ministry of Education, Culture, Sports, Science, and Technology (MEXT) in Japan and Grant-in-Aid for Scientific Research (15K05093, 17H01130) from the Japan Society for Promotion  of Science (JSPS).
For providing high performance computing resources, Computing Research Center, KEK, JLDG on SINET4 of NII, Research Center for Nuclear Physics,Osaka University, Yukawa Institute of Theoretical Physics, Kyoto University, and Information Technology Center, University of Tokyo are acknowledged. 
This work was partly supported by research programs at K-computer of the RIKEN AICS, HPCI Strategic Program of Japanese MEXT, “Priority Issue on Post-K computer” (Elucidation of the Fundamental Laws and Evolution of the Universe) and Joint Institute for Computational Fundamental Sciences (JICFus).
\end{acknowledgments}

%

\end{document}